\documentclass[12pt]{article}
\usepackage{a4wide}
\usepackage{amssymb}
\usepackage{graphicx}
\begin{document}
{\renewcommand{\thefootnote}{\fnsymbol{footnote}}
\hfill   CGPG--03/3-5 \\
\medskip
\hfill gr--qc/0303073\\
\medskip
\begin{center}
{\LARGE  Homogeneous Loop Quantum Cosmology }\\
\vspace{1.5em}
Martin Bojowald\footnote{e-mail address: {\tt bojowald@gravity.psu.edu}}
\\
\vspace{0.5em}
Center for Gravitational Physics and Geometry,\\
The Pennsylvania State
University,\\
104 Davey Lab, University Park, PA 16802, USA\\
\vspace{1.5em}
\end{center}
}

\setcounter{footnote}{0}

\newcommand{\case}[2]{{\textstyle \frac{#1}{#2}}}
\newcommand{\lP}{l_{\mathrm P}}
\newcommand{\HE}{H^{({\rm E})}}
\newcommand{\hatHE}{\hat{H}^{({\rm E})}}

\newcommand{\md}{{\mathrm{d}}}
\newcommand{\id}{\mathop{\mathrm{id}}}
\newcommand{\tr}{\mathop{\mathrm{tr}}}
\newcommand{\sgn}{\mathop{\mathrm{sgn}}}
\newcommand{\semidir}{\mathrel{\mathrm{\times\mkern-3.3mu\protect%
\rule[0.04ex]{0.04em}{1.05ex}\mkern3.3mu\mbox{}}}}

\newcommand*{\R}{{\mathbb R}}
\newcommand*{\N}{{\mathbb N}}
\newcommand*{\Z}{{\mathbb Z}}
\newcommand*{\Q}{{\mathbb Q}}
\newcommand*{\C}{{\mathbb C}}

\begin{abstract}
 Loop quantum cosmological methods are extended to homogeneous models
 in diagonalized form. It is shown that the diagonalization leads to a
 simplification of the volume operator such that its spectrum can be
 determined explicitly. This allows the calculation of composite
 operators, most importantly the Hamiltonian constraint. As an
 application the dynamics of the Bianchi I model is studied and it is
 shown that its loop quantization is free of singularities.
\end{abstract}

\section{Introduction}

In applications so far, quantum gravity has mostly been analyzed in
mini- or midi-superspace models which are obtained by ignoring an
infinite number of degrees of freedom of the full theory. In
particular, as a first approximation of our universe, which is
homogeneous and isotropic at large scales, one often uses a model with
a single gravitational degree of freedom, the scale factor or radius
$a$ of an isotropic universe \cite{DeWitt}. To study more realistic
models, and to see whether results of the highly symmetric
approximation are robust, one has to break some of the symmetries. As
a first step, one can break isotropy while retaining homogeneity. This
provides a large class of models much more general than the isotropic
ones which still have a finite number of degrees of freedom such that
field theoretic complications can be avoided.

The main question for quantum cosmology is the fate of the classical
singularity. In isotropic loop quantum cosmology \cite{IsoCosmo} the
evolution has been shown to be singularity-free \cite{Sing}, and to
lead to additional unexpected quantum effects \cite{DynIn,Inflation}.
To check the robustness of those results one has to see whether they
persist in less symmetric, e.g.\ homogeneous, models. At the same
time, homogeneous models show many different effects in their
classical approach to the singularity which may obtain quantum
modifications. Homogeneous models also provide a hint at the behavior
of the full theory: according to the BKL-scenario \cite{BKL} different
points on a space-like slice decouple from each other when one
approaches the singularity which suggests that it is enough to
understand homogeneous models in order to see the fate of the
classical singularity. Of course, the BKL-scenario itself is not
established and so far it is purely classical. Still, the behavior of
homogeneous models can provide suggestions as to the general approach
to a classical singularity before we are able to study this issue in
the full theory.

Besides the reduction of degrees of freedom, the main simplification
of isotropic loop quantum cosmology as compared to full quantum
geometry is the fact that its volume operator is simple and its
spectrum is known explicitly \cite{cosmoII}. This is essential for
direct computations since the volume operator plays an important role
in constructing more complicated operators such as the Hamiltonian
constraint \cite{QSDI}. If one can deal with the volume operator
easily, one can also obtain an explicit Hamiltonian constraint
equation which then can at least be implemented numerically. This has
been essential in recent physical applications of isotropic loop
quantum cosmology
\cite{InvScale,Sing,DynIn,Ambig,Inflation,Scalar,Closed}. Homogeneous
models \cite{cosmoI} still have a finite number of degrees of freedom,
but their volume operator corresponds to that of the full theory
restricted to a spin network vertex with six incoming edges
\cite{cosmoII}. It is impossible to compute the spectrum of this
operator explicitly since it would require a diagonalization of
arbitrarily large matrices.

However, this volume operator is that of a general homogeneous model
which has six kinematical gravitational degrees of freedom (all
components of a symmetric $3\times3$-matrix which represents the
metric in a given point). In classical and standard quantum
cosmological analyses one usually considers only diagonalized models
with only the three diagonal components as kinematical degrees of
freedom. Because the remaining components can be chosen to be zero by
fixing gauge or symmetry transformations (depending on the model
\cite{AshSam}), diagonalized models are enough for physical applications.
Diagonalizing a homogeneous model is a restriction to a submodel since
some degrees of freedom are ignored. In its physical interpretation
this is different from a symmetry reduction, but the same techniques
(developed in \cite{SymmRed}) apply.

It turns out that this reduction suffices to simplify the volume
operator resulting in an explicitly known spectrum. As in the
isotropic case, this also requires to deal with generalized spin
network states which do not appear as functions on a group
manifold. One can represent those states in a convenient form such
that an explicit analysis of homogeneous models is possible. The main
application in the present paper is the result that the evolution of
homogeneous models is non-singular, which occurs in a way very similar
to that in isotropic models. This is illustrated here in the Bianchi I
model corresponding to a torus action of the symmetry group.

Compared to isotropic models, the absence of a singularity in
homogeneous models relies quite non-trivially on properties of quantum
geometry. In isotropic models it has been seen to be important that
quantum geometry is based on triad variables rather than the
metric. This introduces a sign in the degree of freedom corresponding
to the metric which is necessary to follow the evolution through the
singularity. In homogeneous models we will observe the same effect,
but it could also have been obtained if quantum geometry would be
based on, e.g., the co-triad rather than the densitized triad. We will
see that the removal of homogeneous singularities (contrary to
isotropic singularities) would not happen generically in a theory
based on the co-triad. The reason is that the classical singularity
lies in the interior of a minisuperspace built from the densitized
triad, while it lies at the boundary (at infinity) of a minisuperspace
based on the co-triad.

After demonstrating how the diagonalization of a homogeneous model can
be done in a connection formulation we will implement it at the
quantum level. This requires a reduction of general homogeneous models
along the lines of a symmetry reduction which is carried out
explicitly. In this way we will obtain the kinematical framework whose
structure is very similar to that of isotropic models, and derive the
volume operator and its spectrum. This will allow us to compute the
Hamiltonian constraint equation explicitly and to bring it into the
form of a partial difference equation after transforming to the triad
representation. As an application the classical singularity is shown
to be absent.

\section{Diagonal Bianchi Class A Models}

Bianchi models are obtained by a symmetry reduction of general
relativity with respect to a symmetry group $S$ which acts freely and
transitively on the space manifold $\Sigma\cong S$. Only class A
models which have a symmetry group with structure constants fulfilling
$c^I_{IJ}=0$ are amenable to a Hamiltonian analysis. The action
provides left-invariant 1-forms $\omega_a^I$ which yield the
decomposition $A_a^i=\phi_I^i\omega_a^I$ of an invariant connection
with constant coefficients $\phi_I^i$. While the left-invariant
1-forms are regarded as background structure (provided by the symmetry
group) with respect to which the reduction is being done, the
components $\phi_I^i$ describe the invariant connection in a reduced
model. A corresponding decomposition of the densitized triad is given
by $E_i^a=p_i^IX_I^a$ where $X_I^a$ are densitized left-invariant
vector fields dual to the 1-forms. The symplectic structure of the
reduced model is given by
$\{\phi_I^i,p_j^J\}=\gamma\kappa\delta_j^i\delta_I^J$ where $\gamma$
is the Barbero--Immirzi parameter and $\kappa=8\pi G$ the
gravitational constant. (The role of a possible coordinate volume is
discussed in \cite{Bohr}.)

In this form, the invariant connection has nine independent components
which, after subtracting three gauge degrees of freedom, yields six
gauge-invariant components. This corresponds to the fact that a
homogeneous model is described by $\phi_I^i$ in full generality
without restricting to diagonal metrics. To perform the
diagonalization we rewrite an invariant connection by using the polar
decomposition $\phi=R\psi$ where $R\in SO(3)$ is a rotation matrix and
$\psi$ a symmetric matrix. The symmetric matrix $\psi=\Lambda{\rm
diag}(c_1,c_2,c_3)\Lambda^T$ can then be diagonalized introducing
another rotation matrix $\Lambda\in SO(3)$. Defining $O:=R\Lambda\in
SO(3)$ and redefining $\omega_a^{\prime K}:=\omega_a^IO_I^K$ we obtain
\begin{eqnarray}
 A_a^i &=& \phi_I^i\omega_a^I= R_I^J\psi_J^i\omega_a^I =
 R_I^J\Lambda_J^Kc_{(K)}\Lambda_K^i\omega_a^I\nonumber\\
 &=& c_{(K)}\Lambda_K^i\omega_a^{\prime K}\,. \label{DiagA}
\end{eqnarray}
Now we can clearly see that only three gauge-invariant degrees of
freedom $c_I$ remain, which is just what we need for the
diagonalization. (Note that we absorbed the rotation matrix $O$ into
the left-invariant 1-forms, which can be done in general without
amounting to a reduction of the theory. In a reduced model, however,
we have to fix the 1-forms in the beginning as a background structure,
and then different forms differing by a rotation would lead to
different formulations of a model. This explains that the
diagonalization leads to a reduction of the model, while any
connection can be written in the form (\ref{DiagA}) without
restriction if the $\omega_I$ are regarded as being free. Note also
that the diffeomorphism constraint ${\cal D}_I=c^J_{IK}\phi^i_Jp^K_i=
c^J_{IK}c_{(J)}p^{(K)}\delta^K_J$ vanishes identically in the
diagonalized class A models whereas it generates inner automorphisms
of the symmetry group in non-diagonal models.)

In fact, the corresponding decomposition 
\begin{equation}\label{DiagE}
 E_i^a=p^{(K)}\Lambda_i^KX_K^{\prime a}
\end{equation}
of the triad shows that the metric is diagonal. We have to allow
negative values for the coefficients $p^I$ since the triad can have
two different orientations while the orientation of $\Lambda\in SO(3)$
is fixed.  The $SO(3)$-matrix $\Lambda$ contains purely gauge degrees
of freedom and plays the same role as the matrix $\Lambda$ in the
isotropic decomposition \cite{IsoCosmo}, even though it arose in a
different way. The advantage of the diagonalization in the connection
formulation is that gauge-invariant degrees of freedom in $c_I$ and
gauge degrees of freedom in $\Lambda$ are strictly separate (except
for simple residual gauge transformations acting on the $c_I$, and
correspondingly on $p^I$, which will be discussed in
Sec.~\ref{DiagSpin}). As in the isotropic case, this will lead to a
simplification of the volume operator since essentially
$SU(2)$-calculations can be reduced to $U(1)$-calculations. In the
diagonal components the symplectic structure is
\begin{equation} \label{symp}
 \{c_I,p^J\}=\gamma\kappa\delta_I^J\,.
\end{equation}

Also the co-triad $e_a^i=a_I^i\omega_a^I=a_{(I)}\Lambda_I^i\omega_a^I$
has the diagonal form with arbitrary real parameters $a_I\in\R$. The
relation between the triad and co-triad components follows from
$E_i^a=|\det e|(e_a^i)^{-1}$ and is given by
\begin{equation} \label{pa}
 p^1=|a_2a_3|\sgn(a_1)\quad,\quad p^2=|a_1a_3|\sgn(a_2) \quad,\quad
 p^3=|a_1a_2|\sgn(a_3)\,.
\end{equation}
In these variables the volume is given by
\[
 V=\sqrt{\left|\case{1}{6}
     \epsilon^{ijk}\epsilon_{IJK}p_i^Ip_j^Jp_k^K\right|}
 =\sqrt{|p^1p^2p^3|} =|a_1a_2a_3|\,.
\]

For the geometrical interpretation of the connection components we
need the homogeneous spin connection
$\Gamma_a^i=\Gamma_{(I)}\Lambda_I^i\omega_a^I$. From the general
formula \cite{AsFlat}
\[
 \Gamma_a^i=-\case{1}{2}\epsilon^{ijk}e_j^b (2\partial_{[a} e_{b]}^k+
 e_k^ce_a^l \partial_c e_b^l)
\]
we derive
\begin{eqnarray} \label{SpinConn}
 \Gamma_I &=& \case{1}{2}\left(\frac{a_J}{a_K}n^J+ \frac{a_K}{a_J}n^K-
   \frac{a_I^2}{a_Ja_K}n^I\right)\\
  &=& \case{1}{2}\left(\frac{p^K}{p^J}n^J+ \frac{p^J}{p^K}n^K-
   \frac{p^Jp^K}{(p^I)^2}n^I\right)
  \qquad\mbox{for $\epsilon_{IJK}=1$}\nonumber 
\end{eqnarray}
where the numbers $n^I$ specify the Bianchi model via
$c^I_{JK}=\epsilon_{JKL}n^{IL}=\epsilon_{JK}^In^{(I)}$.
Only the Bianchi I model has vanishing spin connection; otherwise
$\Gamma_I$ depends on the triad components, contrary to isotropic
models where spin connection components are constant.

For diagonal models the Euclidean part of the Hamiltonian
constraint \cite{AshVar,AshVarReell} reduces to
\begin{eqnarray}
 \HE &=& -\kappa^{-1}\det(e_I^i)^{-1}\epsilon_{ijk}
  F^i_{IJ}E^I_jE^J_k\nonumber\\
 &=& \kappa^{-1}\det(a_I^i)^{-1}
  (\epsilon_{ijk}c^K_{IJ} \phi^i_Kp^I_jp^J_k-
  \phi^j_I\phi^k_Jp^I_jp^J_k+ \phi^k_I\phi^j_Jp^I_jp^J_k)\nonumber\\
 &=& 2\kappa^{-1}\left((n^1c_1-c_2c_3)a_1+(n^2c_2-c_1c_3)a_2+
   (n^3c_3-c_1c_2)a_3\right)\,.
  \label{HE}
\end{eqnarray}
The full constraint $H=-\HE+P$ with
\[
 P = -2(1+\gamma^2)\kappa^{-1}\det(e_I^i)^{-1} K_{[I}^iK_{J]}^j
 E_i^IE_j^J\,,
\]
using the extrinsic curvature
$K_I^i=\gamma^{-1}(\Gamma_I^i-\phi_I^i)$, is
\begin{eqnarray} \label{H}
 H &=& 2\kappa^{-1}\left(\left((c_2\Gamma_3+c_3\Gamma_2-\Gamma_2\Gamma_3)
     (1+\gamma^{-2})- n^1c_1-\gamma^{-2}c_2c_3\right)a_1\right.\nonumber\\
  &&+\left((c_1\Gamma_3+c_3\Gamma_1-\Gamma_1\Gamma_3)
     (1+\gamma^{-2})- n^2c_2-\gamma^{-2}c_1c_3\right)a_2\nonumber\\
  &&\left.+\left((c_1\Gamma_2+c_2\Gamma_1-\Gamma_1\Gamma_2)
     (1+\gamma^{-2})- n^3c_3-\gamma^{-2}c_1c_2\right)a_3\right)\,.
\end{eqnarray}
In terms of extrinsic curvature components
$K_I=\frac{1}{2}\dot{a}_I=\gamma^{-1}(\Gamma_I-c_I)$, where the dot
indicates a derivative with respect to proper time $t$ corresponding
to lapse $N=1$, it reads
\begin{eqnarray} \label{Ha}
 H &=& 2\kappa^{-1}\left((\Gamma_2\Gamma_3-n^1\Gamma_1)a_1+
   (\Gamma_1\Gamma_3-n^2\Gamma_2)a_2
   +(\Gamma_1\Gamma_2-n^3\Gamma_3)a_3\right.\nonumber\\
  &&\left. -\case{1}{4}(a_1\dot{a}_2\dot{a}_3+ a_2\dot{a}_1\dot{a}_3+
   a_3\dot{a}_1\dot{a}_2)\right)
\end{eqnarray}
which  yields   the    Hamiltonian   constraint   equation   $H+H_{\rm
  matter}(a_1,a_2,a_3)=0$ as  a differential equation for the co-triad
components $a_I$.

As special cases we obtain isotropic models if we set $a_1=a_2=a_3=:a$
such that
\[
 H+H_{\rm matter}(a)= -\case{3}{2}\kappa^{-1}(\dot{a}^2+2\Gamma)a+
 H_{\rm matter}(a)=0
\]
where $\Gamma=0$ for the flat model and $\Gamma=\frac{1}{2}$ for the
model with positive spatial curvature. In this case, the Hamiltonian
constraint equation is the Friedmann equation.

An example for an anisotropic model is the vacuum Bianchi I model where
$n^I=0$ and $H_{\rm matter}=0$. Therefore, the constraint equation
reads
\[
 a_1\dot{a}_2\dot{a}_3+ a_2\dot{a}_1\dot{a}_3+
   a_3\dot{a}_1\dot{a}_2=0\,.
\]
A simple solution is the one where the $a_I$ are constant such that
all terms in the constraint equation vanish separately; this solution
is Minkowski space.  All other solutions are given by the well-known
Kasner solutions \cite{Kasner} $a_I\propto t^{\alpha_I}$ where the
constant exponents $\alpha_I$ have to satisfy
$\sum_I\alpha_I=1=\sum_I\alpha_I^2$. These relations can be fulfilled
only for $-1<\alpha_I\leq1$ and, furthermore, one coefficient
$\alpha_I$ has to be negative while the other two are positive. This
leads to an evolution of the form of Kasner epochs where one direction
of space is contracting while the other two are expanding. The total
volume $V=a_1a_2a_3\propto t$ always increases monotonically. In these
co-triad (or metric) variables $a_I$ the structure of the singularity
at $t=0$ is quite complicated because one metric component diverges.
This implies that the classical singularity lies at the boundary of
minisuperspace.

In the densitized triad variables $p^I$ on which the loop quantum
cosmological description is based the situation turns out to be
different. Using the relation between co-triad and densitized triad
components and the relations for the coefficients $\alpha_I$ it
follows that $p^I\propto t^{1-\alpha_I}$ which has always a positive
exponent. Thus, the triad components are always increasing and the
singularity lies at $p^I=0$ which is an interior point of
minisuperspace. This will turn out to be crucial for the generic
removal of the classical singularity.

Other Bianchi models have a non-vanishing potential given by the spin
connection components. This leads to possible changes in the
expansion/contraction behavior of the co-triad components with
intermediate regimes approximately described by Kasner epochs. Such a
succession of different epochs can even be chaotic in certain models
which renders the structure of the classical singularity even more
complicated. In densitized triad variables, the situation looks again
simpler since all components decrease during any Kasner epoch
approaching the singularity. Only the component which decreases the
fastest changes during the succession of different epochs.

\section{Quantum Theory of Diagonal Bianchi Class A Models}

According to the general scheme of a symmetry reduction for
diffeomorphism invariant quantum field theories of connections
\cite{SymmRed}, symmetric states are distributional states which are
supported only on invariant connections with respect to the given
symmetry group. For homogeneous models those states are in one-to-one
correspondence with functions on the group manifold $SU(2)^3$ which
can be visualized as spin network states associated with a graph
containing a single vertex and three closed edges intersecting in the
$6$-vertex. For the diagonalization we have to perform an additional
reduction to diagonal homogeneous connections (\ref{DiagA}). Even
though this is not a symmetry reduction, the same techniques
apply. Consequently, diagonal homogeneous states are functions on the
manifold
\begin{equation}\label{Conf}
 \{(\exp(c_1\Lambda_1^i\tau_i),\exp(c_2\Lambda_2^i\tau_i),
 \exp(c_3\Lambda_3^i\tau_i)):c_I\in\R,\Lambda\in SO(3)\}\subset SU(2)^3
\end{equation}
which is obtained by building (point) holonomies out of the connection
components (this issue is discussed more carefully in
\cite{Bohr}). The diagonal homogeneous configuration space is by
definition a submanifold of the general homogeneous configuration
space $SU(2)^3$, but as in the isotropic case \cite{cosmoII} one can
easily check that it is not a subgroup. This implies that the
Peter--Weyl theorem cannot be used to find a generating set of
functions, and states have to be expected not to be ordinary spin
network states. In the isotropic case all possible functions of the
lone gauge invariant connection component $c$ were allowed as gauge
invariant states, not just even functions as would be expected by a
naive reduction (corresponding to characters on a single copy of
$SU(2)$).

\subsection{Diagonal Homogeneous Spin Network States and Basic Operators}
\label{DiagSpin}

To derive all possible states for a given model one has to restrict
the already known homogeneous states (or in general all states of the
full theory) to the subspace of reduced connections, find an
independent set, and complete it with respect to the measure on the
space of reduced connections. The resulting set can be orthonormalized
to derive a generalized spin network basis. In most cases this
procedure can be simplified if a decomposition of reduced connections
into pure gauge degrees of freedom and remaining components is
known. Then, gauge invariant states can be found to be arbitrary
functions of the remaining components subject possibly to residual
gauge transformations which are usually discrete and easy to find. For
instance, in the isotropic case an invariant connection has the form
$A_a^i=c\Lambda_I^i\omega_a^I$ where the components $\Lambda_I^i$ are
pure gauge. There are no residual gauge transformations on $c$ (note
that changing the sign of $c$ is equivalent to changing the
orientation of $\Lambda$, which is not a gauge transformation) such
that gauge invariant states can be arbitrary functions of $c$ without
restrictions. This basic fact turns out to have important consequences
even for the dynamics and physical applications: it implies that
states are labeled by an integer quantum number $n$ rather than a
positive integer $2j+1$ as would be expected for an $SU(2)$-theory. In
the end, this leads to the removal of classical singularities in
isotropic models.

We can now perform this analysis for diagonal homogeneous connections
(\ref{DiagA}) which also allow a decomposition into pure gauge degrees
of freedom $\Lambda_I^i$ and remaining components $c_I$. We start with
noting the measure on the space of gauge invariant diagonal
homogeneous connections as a quotient of (\ref{Conf}). Since the
relations between the three copies of $SU(2)$ implied by the
diagonalization only affect the gauge degrees of freedom
$\Lambda_I^i$, the gauge invariant measure is a direct product of
three copies of the Haar measure on $SU(2)$. Therefore, we
use
\begin{equation}
 \md\mu(c_1,c_2,c_3)= (2\pi)^{-3}
 \sin^2(\case{1}{2}c_1)\sin^2(\case{1}{2}c_2)\sin^2(\case{1}{2}c_3)
 \md c_1\md c_2\md c_3\,.
\end{equation}
Gauge invariant states have to be functions of $c_I$, but this time
there are residual gauge transformations: it can be seen from
(\ref{Conf}) that one can change the sign of two components $c_I$
simultaneously while keeping the third one fixed (e.g., a gauge
transformation with $\exp(\pi\Lambda_3^i\tau_i)$ changes the sign of
$c_1$ and $c_2$). Those are the only residual gauge transformations
such that gauge invariant states have to be invariant under changing
two signs of the $c_I$.

To write down a basis we use the isotropic states given by
\begin{equation}\label{chi}
 \chi_j(c)=\frac{\sin(j+\case{1}{2})c}{\sin\case{1}{2}c}\quad,\quad
 j\in\case{1}{2}{\mathbb N}_0
\end{equation}
together with $\zeta_{-\frac{1}{2}}(c)= (\sqrt{2} \sin\case{1}{2}c)^{-1}$
and
\begin{equation}\label{zeta}
 \zeta_j(c)=\frac{\cos(j+\case{1}{2})c}{\sin\case{1}{2}c}\quad,\quad
 j\in\case{1}{2}{\mathbb N}_0
\end{equation}
which are orthonormalized with respect to the measure
$\md\mu(c)=(2\pi)^{-1}\sin^2(\case{1}{2}c)\md c$. Thus, all
orthonormal states of the diagonal homogeneous model can be written as
$f_{j_1}(c_1)g_{j_2}(c_2)h_{j_3}(c_3)$ where $f$, $g$, $h$ can be
$\chi$ or $\zeta$ and $j_I$ is a non-negative half-integer for $\chi$
and a half-integer larger than $-1$ for $\zeta$. However, only two
different combinations are invariant under the residual gauge
transformations, namely
\begin{equation}\label{states}
 \chi_{j_1,j_2,j_3}(c_1,c_2,c_3):=
 \chi_{j_1}(c_1)\chi_{j_2}(c_2)\chi_{j_3}(c_3)
 \quad,\quad \zeta_{j_1,j_2,j_3}(c_1,c_2,c_3):=
 \zeta_{j_1}(c_1)\zeta_{j_2}(c_2)\zeta_{j_3}(c_3)
\end{equation}
where the $j_I$ can be $-\frac{1}{2}$ only for $\zeta$. As in the
isotropic case, we obtain twice as many states as expected from
positive spin labels. Such a doubling of the number of states has been
seen to be necessary in the isotropic model because it yields
eigenvalues for the triad operator corresponding to the two different
triad orientations. To verify this in the diagonal homogeneous model
we have to consider the quantization of the triad components $p^I$.

As in \cite{cosmoII}, invariant vector field operators which
quantize triad components in the full homogeneous model reduce to
the self-adjoint derivative operators
\begin{equation}
 \hat{p}^I=-i\gamma\lP^2\left(\frac{\partial}{\partial c_I}+
   \case{1}{2}\cot(\case{1}{2}c_I)\right)\,.
\end{equation}
These operators, however, do not leave the space of states
(\ref{states}) invariant corresponding to the fact that the triad
components $p^I$ are not gauge invariant. As with connection
components, we can change two signs of the $p^I$ simultaneously by a
gauge transformation. Thus, only $|p^I|$ and $\sgn(p^1p^2p^3)$ are
gauge invariant, the latter quantity being the orientation. Still, for
calculations it is convenient to work with the non-gauge invariant
$p^I$-operators and to use their eigenstates which have to be
non-gauge invariant. These eigenstates can easily be seen to be
\begin{equation}\label{n}
 |n_1,n_2,n_3\rangle:= |n_1\rangle\otimes |n_2\rangle\otimes
 |n_3\rangle
\end{equation}
where
\[
 \langle c|n\rangle=\frac{\exp(\case{1}{2}inc)}{\sqrt{2}\sin(\case{1}{2}c)}
\]
are the isotropic $p$-eigenstates in the connection
representation. Using Euler's formula for the exponential shows that
the eigenstates (\ref{n}) are not linear combinations of the gauge
invariant states (\ref{states}) only. A gauge invariant state can be
expanded in the states (\ref{n}) as
\begin{equation}\label{TriadRep}
 |s\rangle=\sum_{n_1,n_2,n_3}s_{n_1,n_2,n_3}|n_1,n_2,n_3\rangle
\end{equation}
where the coefficients $s_{n_1,n_2,n_3}$, which represent the state in
the triad representation, have to fulfill
\begin{equation}
 s_{n_1,n_2,n_3}=s_{-n_1,-n_2,n_3}=s_{n_1,-n_2,-n_3}=
 s_{-n_1,n_2,-n_3}\,.
\end{equation}

The set of basic operators is completed by quantizing the connection
components which has to be done via (point) holonomies
$h_I=\exp(c_I\Lambda_I^i\tau_i)=
\cos(\case{1}{2}c_I)+2\Lambda_I^i\tau_i\sin(\case{1}{2}c_I)$ as
multiplication operators. The matrix $\Lambda$ is not gauge invariant,
as a multiplication operator, and does not even leave the space of
states (\ref{n}) invariant. Nevertheless, we do not need to extend our
state space once more because, as shown in \cite{InvScale}, $\Lambda$
drops out of gauge invariant operators even when holonomies are used
in intermediate steps (as in commutators with the volume operator). We
then need only the action of $\cos(\case{1}{2}c_I)$ and
$\sin(\case{1}{2}c_I)$ which can readily be seen to be
\begin{eqnarray}
 \cos(\case{1}{2}c_1) |n_1,n_2,n_3\rangle &=&
 \case{1}{2}(|n_1+1,n_2,n_3\rangle+|n_1-1,n_2,n_3\rangle)\\ 
 \sin(\case{1}{2}c_1) |n_1,n_2,n_3\rangle &=&
 -\case{1}{2}i(|n_1+1,n_2,n_3\rangle- |n_1-1,n_2,n_3\rangle)
\end{eqnarray}
and correspondingly for $c_2$ and $c_3$.

\subsection{Gauge Invariant Triad Operators}

As already discussed, the triad components $p^I$ are not completely
gauge invariant, but only the functions $|p^I|$ and $\sgn(p^1p^2p^3)$
are. The corresponding operators must leave the gauge invariant states
(\ref{n}) invariant. In fact, we have
\[
 |\hat{p}^I|\chi_{j_1,j_2,j_3}=\gamma\lP^2(j_I+\case{1}{2})
 \chi_{j_1,j_2,j_3} \quad,\quad
 |\hat{p}^I|\zeta_{j_1,j_2,j_3}=\gamma\lP^2(j_I+\case{1}{2})
 \zeta_{j_1,j_2,j_3}
\]
and
\begin{eqnarray*}
 \hat{p}^1\hat{p}^2\hat{p}^3\chi_{j_1,j_2,j_3} &=& i(\gamma\lP^2)^3
 (j_1+\case{1}{2}) (j_2+\case{1}{2}) (j_3+\case{1}{2})
 \zeta_{j_1,j_2,j_3} \\
 \hat{p}^1\hat{p}^2\hat{p}^3\zeta_{j_1,j_2,j_3} &=& -i(\gamma\lP^2)^3 
 (j_1+\case{1}{2}) (j_2+\case{1}{2}) (j_3+\case{1}{2})
 \chi_{j_1,j_2,j_3}\,.
\end{eqnarray*}
Simultaneous eigenstates of all gauge invariant triad operators are
given by
\[
 2^{-\frac{1}{2}}(\zeta_{j_1,j_2,j_3}\pm i\chi_{j_1,j_2,j_3})\,.
\]

In terms of the states (\ref{n}) the action reads
\[
 |\hat{p}^I|\;|n_1,n_2,n_3\rangle=
 \case{1}{2}\gamma\lP^2|n_I|\;|n_1,n_2,n_3\rangle
\]
and
\[
 \sgn(\hat{p}^1\hat{p}^2\hat{p}^3)|n_1,n_2,n_3\rangle=
 \sgn(n_1n_2n_3)|n_1,n_2,n_3\rangle\,.
\]
Gauge invariant simultaneous eigenstates are
\begin{eqnarray}
 && \case{1}{2}(|n_1,n_2,n_3\rangle+ |-n_1,-n_2,n_3\rangle+
 |-n_1,n_2,-n_3\rangle+ |n_1,-n_2,-n_3\rangle)\\
 && \quad= 2^{-\frac{1}{2}}
 \left(\zeta_{\frac{1}{2}(|n_1|-1), \frac{1}{2}(|n_2|-1),
   \frac{1}{2}(|n_3|-1)}- i\sgn(n_1n_2n_3) \chi_{\frac{1}{2}(|n_1|-1),
   \frac{1}{2}(|n_2|-1), \frac{1}{2}(|n_3|-1)}\right)\,.\nonumber
\end{eqnarray}

From the triad operators we can immediately build the volume operator
$\hat{V}=\sqrt{|\hat{p}^1\hat{p}^2\hat{p}^3|}$ (it can also be derived
from the non-diagonal homogeneous volume operator \cite{cosmoII},
which itself descends from the full one) with eigenvalues
\begin{equation}\label{Vol}
 \hat{V}|n_1,n_2,n_3\rangle= (\case{1}{2}\gamma\lP^2)^{\frac{3}{2}}
 \sqrt{|n_1n_2n_3|}\; |n_1,n_2,n_3\rangle=: V_{n_1,n_2,n_3}
 |n_1,n_2,n_3\rangle\,.
\end{equation}
To avoid confusion we note that the isotropic states $|n\rangle$ of
\cite{IsoCosmo} are not to be identified with states
$|n,n,n\rangle$. Rather, according to the general scheme of a symmetry
reduction an isotropic state would correspond to a distributional
state in a homogeneous model and thus be non-normalizable. Therefore,
the isotropic volume eigenvalues need not be given by $V_{n,n,n}$.

Using the basic operators we can construct more complicated ones like
inverse metric operators along the lines of
\cite{QSDI,QSDV,InvScale}. Since there are no essential differences to
the calculations in the isotropic model, we will not repeat details
here. Those operators are needed when one quantizes the spin
connection (\ref{SpinConn}) or matter Hamiltonians.

\section{Dynamics of the Bianchi I Model}

So far we considered only kinematical aspects, which are independent of
the particular type of Bianchi model. Specific properties only enter
via the Hamiltonian constraint (\ref{H}) (or the spin connection)
which explicitly contains the structure constants of the symmetry
group. Bianchi models usually have a non-vanishing spin connection
which is necessary to yield non-zero intrinsic curvature. In the full
theory such a statement would be meaningless because the spin
connection does not have an invariant meaning. More precisely, it can
be made arbitrarily small locally by choosing appropriate coordinates.
In symmetric models, however, we are restricted to coordinate choices
which respect the symmetry such that the spin connection does have a
coordinate independent meaning in homogeneous models (or certain
components in a less symmetric model). This also implies that it is
impossible to ignore the effects of a possibly large spin connection
whenever there is non-zero intrinsic curvature. As a consequence, even
in classical regimes where, e.g., the extrinsic curvature is small,
there can still be large intrinsic curvature contributions. Since this
effect is a consequence of the symmetry and is not present in the full
theory, it has to be dealt with appropriately. For instance, it
affects the quantization of the Hamiltonian constraint (which has not
been taken properly into account for the general quantization of the
Hamiltonian constraints in homogeneous models presented in
\cite{cosmoIII}) and the interpretation of the semiclassical limit;
this is discussed in more detail in \cite{Spin}.

The lone exception to the preceding statements is the Bianchi I model
which has vanishing intrinsic curvature and where the spin connection
is always zero. Consequently, its quantization is simpler and also
closer to that of the full theory \cite{QSDI}. In the present paper we
restrict our attention to this model and in particular to the issue of
the classical singularity displayed in the Kasner behavior.

\subsection{Hamiltonian Constraint}

From \cite{cosmoIII} we obtain the quantized Euclidean part of the
Hamiltonian constraint in the form
\[
 \hatHE=4i(\gamma\kappa\lP^2)^{-1} \sum_{IJK}\epsilon^{IJK}
 \tr(h_Ih_Jh_I^{-1}h_J^{-1}h_K[h_K^{-1},\hat{V}])\,.
\]
A further simplification of the Bianchi I model is that the Lorentzian
constraint is related to the Euclidean part simply by
$H=\gamma^{-2}\HE$ which we make use of here. Alternatively, one could
derive the Lorentzian constraint as in the full theory \cite{QSDI} by
expressing the extrinsic curvature as a commutator of the Euclidean
part and the volume. Since this has been done in detail in the
isotropic model without yielding essential differences to the simpler
approach adopted here \cite{IsoCosmo}, we will not perform the full
analysis. In other Bianchi models the relation between the Lorentzian
constraint and the Euclidean part is more complicated and one would
have to quantize the additional terms (containing the spin connection)
by hand, introducing more ambiguities \cite{Spin}.

Using diagonal (point) holonomies $h_I=\cos(\case{1}{2}c_I)+
2\Lambda_I^i\tau_i\sin(\case{1}{2}c_I)$ in
$\hat{H}:=\gamma^{-2}\hatHE$, we obtain
\begin{eqnarray*}
 \hat{H} &=& 32i\gamma^{-3}\kappa^{-1}\lP^{-2} \left(\sin(\case{1}{2}c_1)
   \cos(\case{1}{2}c_1) \sin(\case{1}{2}c_2) \cos(\case{1}{2}c_2)\right.\\
 &&\qquad \times  (\sin(\case{1}{2}c_3) \hat{V} \cos(\case{1}{2}c_3)-
   \cos(\case{1}{2}c_3) \hat{V} \sin(\case{1}{2}c_3))\\
 && +\sin(\case{1}{2}c_3) \cos(\case{1}{2}c_3) \sin(\case{1}{2}c_1)
 \cos(\case{1}{2}c_1) (\sin(\case{1}{2}c_2) \hat{V} \cos(\case{1}{2}c_2)-
   \cos(\case{1}{2}c_2) \hat{V} \sin(\case{1}{2}c_2))\\
 && +\left.\sin(\case{1}{2}c_2) \cos(\case{1}{2}c_2) \sin(\case{1}{2}c_3)
 \cos(\case{1}{2}c_3) (\sin(\case{1}{2}c_1) \hat{V}
 \cos(\case{1}{2}c_1)- \cos(\case{1}{2}c_1) \hat{V}
 \sin(\case{1}{2}c_1))\right)
\end{eqnarray*}
with action
\begin{eqnarray*}
 \hat{H}|n_1,n_2,n_3\rangle &=& \gamma^{-3}\kappa^{-1}\lP^{-2}
 \left((V_{n_1,n_2,n_3+1}-V_{n_1,n_2,n_3-1})
   (|n_1+2,n_2+2,n_3\rangle\right.\\
 &&\quad-|n_1-2,n_2+2,n_3\rangle- |n_1+2,n_2-2,n_3\rangle+
   |n_1-2,n_2-2,n_3\rangle) \\
 &&+(V_{n_1,n_2+1,n_3}-V_{n_1,n_2-1,n_3}) (|n_1+2,n_2,n_3+2\rangle\\
 &&\quad- |n_1-2,n_2,n_3+2\rangle- |n_1+2,n_2,n_3-2\rangle+
   |n_1-2,n_2,n_3-2\rangle)\\
 &&+(V_{n_1+1,n_2,n_3}-V_{n_1-1,n_2,n_3}) (|n_1,n_2+2,n_3+2\rangle\\
 &&\quad-\left. |n_1,n_2+2,n_3-2\rangle- |n_1,n_2-2,n_3+2\rangle+
   |n_1,n_2-2,n_3-2\rangle)\right)\,.
\end{eqnarray*}

To derive an interpretation of the constraint equation
$(\hat{H}+\hat{H}_{\rm matter})|s\rangle=0$ as an evolution equation,
we transform to the triad representation (\ref{TriadRep}) where we
have
\begin{eqnarray}
(\hat{H}s)_{n_1,n_2,n_3} &=&
\gamma^{-3}\kappa^{-1}\lP^{-2} \left( 
   (V_{n_1-2,n_2-2,n_3+1}-V_{n_1-2,n_2-2,n_3-1})
   s_{n_1-2,n_2-2,n_3}\right.\nonumber\\
  &&- (V_{n_1+2,n_2-2,n_3+1}-V_{n_1+2,n_2-2,n_3-1})
  s_{n_1+2,n_2-2,n_3}\nonumber\\
  &&-  (V_{n_1-2,n_2+2,n_3+1}-V_{n_1-2,n_2+2,n_3-1})
   s_{n_1-2,n_2+2,n_3}\nonumber\\
 &&+ (V_{n_1+2,n_2+2,n_3+1}-V_{n_1+2,n_2+2,n_3-1})
 s_{n_1+2,n_2+2,n_3}\nonumber\\
 &&+(V_{n_1-2,n_2+1,n_3-2}-V_{n_1-2,n_2-1,n_3-2})
   s_{n_1-2,n_2,n_3-2} \nonumber\\
 && - (V_{n_1+2,n_2+1,n_3-2}-V_{n_1+2,n_2-1,n_3-2})
  s_{n_1+2,n_2,n_3-2}\nonumber\\
 &&- (V_{n_1-2,n_2+1,n_3+2}-V_{n_1-2,n_2-1,n_3+2})
   s_{n_1-2,n_2,n_3+2}\nonumber\\
 &&+ (V_{n_1+2,n_2+1,n_3+2}-V_{n_1+2,n_2-1,n_3+2})
 s_{n_1+2,n_2,n_3+2}\nonumber\\
 &&+(V_{n_1+1,n_2-2,n_3-2}-V_{n_1-1,n_2-2,n_3-2})
   s_{n_1,n_2-2,n_3-2}\nonumber\\
 && - (V_{n_1+1,n_2+2,n_3-2}-V_{n_1-1,n_2+2,n_3-2})
  s_{n_1,n_2+2,n_3-2}\nonumber\\
 &&- (V_{n_1+1,n_2-2,n_3+2}-V_{n_1-1,n_2-2,n_3+2})
   s_{n_1,n_2-2,n_3+2}\nonumber\\
  &&+ \left.(V_{n_1+1,n_2+2,n_3+2}-V_{n_1-1,n_2+2,n_3+2})
 s_{n_1,n_2+2,n_3+2}\right)\,.
\end{eqnarray}

Using the volume eigenvalues (\ref{Vol}), we obtain the evolution
equation
\begin{eqnarray} \label{Evolve}
 && |n_3+2|^{\frac{1}{2}}
   \left(\left(|n_2+1|^{\frac{1}{2}}-|n_2-1|^{\frac{1}{2}}\right)
   \left(|n_1+2|^{\frac{1}{2}}s_{n_1+2,n_2,n_3+2}-
   |n_1-2|^{\frac{1}{2}}s_{n_1-2,n_2,n_3+2}\right)\right.\nonumber\\
 &&\qquad+\left.\left(|n_1+1|^{\frac{1}{2}}-|n_1-1|^{\frac{1}{2}}\right)
   \left(|n_2+2|^{\frac{1}{2}}s_{n_1,n_2+2,n_3+2}-
   |n_2-2|^{\frac{1}{2}}s_{n_1,n_2-2,n_3+2}\right)\right)\nonumber\\
 &&+\left(|n_3+1|^{\frac{1}{2}}-|n_3-1|^{\frac{1}{2}}\right)\left(
   |n_2-2|^{\frac{1}{2}}\left(|n_1-2|^{\frac{1}{2}} s_{n_1-2,n_2-2,n_3}-
   |n_1+2|^{\frac{1}{2}}s_{n_1+2,n_2-2,n_3}\right)\right.\nonumber\\
 &&\qquad+ \left.|n_2+2|^{\frac{1}{2}} \left(|n_1+2|^{\frac{1}{2}}
   s_{n_1+2,n_2+2,n_3}- 
  |n_1-2|^{\frac{1}{2}} s_{n_1-2,n_2+2,n_3}\right)\right)\nonumber\\
 && +|n_3-2|^{\frac{1}{2}}
   \left(\left(|n_2+1|^{\frac{1}{2}}-|n_2-1|^{\frac{1}{2}}\right)
   \left(|n_1-2|^{\frac{1}{2}}s_{n_1-2,n_2,n_3-2}-
   |n_1+2|^{\frac{1}{2}}s_{n_1+2,n_2,n_3-2}\right)\right.\nonumber\\
 &&\qquad+\left.\left(|n_1+1|^{\frac{1}{2}}-|n_1-1|^{\frac{1}{2}}\right)
   \left(|n_2-2|^{\frac{1}{2}}s_{n_1,n_2-2,n_3-2}-
   |n_2+2|^{\frac{1}{2}}s_{n_1,n_2+2,n_3-2}\right)\right)\nonumber\\
 &=& -2\kappa (\case{1}{2}\gamma\lP^2)^{-\frac{1}{2}} \hat{H}_{\rm
   matter}(n_1,n_2,n_3) s_{n_1,n_2,n_3}
\end{eqnarray}
where $\hat{H}_{\rm matter}$ can be any matter Hamiltonian which
depends on the triad components $n_I$ and acts on the wave function
$s_{n_1,n_2,n_3}$ (we suppressed the dependence of the wave function
on matter fields). Equation (\ref{Evolve}) is a partial difference
equation for the wave function in the triad representation. For a
dynamical interpretation we have to choose an internal time which in
the case of the Bianchi I model can easily be done as one of the
diagonal components of the triad, e.g., $n_3$. This is justified
because classically $p^3$ is always monotonically expanding or
contracting. With this internal time the constraint equation has the
form of an evolution equation in a discrete time, as expected on
general grounds \cite{cosmoIV}.

\subsection{Absence of a Singularity}

The classical Kasner behavior suggests that the approach to the
singularity of the Bianchi I model can well be described with the
internal time $p^3$ (or any other diagonal triad component). In the
loop quantization this leads to the discrete internal time $n_3$ which
we already chose to interpret Equation (\ref{Evolve}) as a time
evolution equation for the wave function $s_{n_1,n_2,n_3}$. The other
labels $n_1$ and $n_2$ can be regarded as independent degrees of
freedom evolving in the time $n_3$.  While the time variable $n_3$ is
unrestricted in the range from $-\infty$ to $+\infty$, for independent
values of the wave function we have to consider only non-negative
$n_1,n_2\geq 0$. Remaining values of the wave function are then given
by the relations
\begin{equation}\label{Ident}
 s_{-n_1,n_2,n_3}=s_{n_1,n_2,-n_3} \quad,\quad
 s_{n_1,-n_2,n_3}=s_{n_1,n_2,-n_3}
\end{equation}
which follow from gauge invariance. In other words, $s$ can be seen as
a function on a discrete space $\{(n_1,n_2,n_3):
n_1,n_2\in\N_0,n_3\in\Z\}$ subject to the identifications
$(n_1,0,n_3)\equiv(n_1,0,-n_3)$ and
$(0,n_2,n_3)\equiv(0,n_2,-n_3)$. Certainly, any one of the discrete
coordinates can be chosen to take negative values; however, it is most
convenient to choose that one which is being used as internal time. As
in isotropic loop quantum cosmology, the difference to the classical
and Wheeler--DeWitt approach is that there is a ``doubling'' of
minisuperspace which comes from the two possible orientations of a
triad. This can be obtained in any triad formulation, but the
non-trivial fact is whether or not the evolution can be extended
meaningfully and uniquely through the boundary corresponding to the
classical singularity. 

To study this question we use the evolution equation
(\ref{Evolve}). We will interpret this equation as an evolution equation in
the internal time $n_3$, but it is important to note that the presence
of a singularity is independent of the choice of internal time. The
question is whether or not the wave function can be extended uniquely
throughout all of minisuperspace even crossing singular
hypersurfaces. For an intuitive picture one usually selects a function
on minisuperspace, most often the volume, as a time coordinate and
views solutions in a semiclassical regime as representing wave packets
evolving in this internal time. But this is only necessary for an
intuitive interpretation, not for a proof of absence of
singularities. In particular, one can choose a time coordinate
different from the conventional volume. Here we use the integer $n_3$,
corresponding to the triad component $p^3$, because it leads to a
difference equation, while the volume does not have equally spaced
eigenvalues and thus does not give a simple evolution
equation. Physical results, e.g.\ expectation values of observables,
can always be translated between different internal time pictures.

As a recurrence relation, equation (\ref{Evolve}) does not only
require initial conditions at slices of constant $n_3$ but also
boundary data for small values of $n_1$ and $n_2$. We will first
assume that the correct amount of boundary conditions is given such
that we can solve the evolution equation for positive $n_3$ and
negative $n_3$, respectively. The issue of initial and boundary
conditions will be discussed later.

Starting with initial data at large positive times $n_3$ and evolving
backwards, equation (\ref{Evolve}) can be used as a recurrence
relation as long as the coefficient $\sqrt{|n_3-2|}$ of the lowest
order part, which contains the wave function with label $n_3-2$, is
non-zero. This is the case only until $n_3=2$ such that our initial
and boundary data uniquely fix the wave function for all $n_3>0$. The
values at singular states, identified with $n_3=0$ or $n_1=0$ or
$n_2=0$, cannot be determined which looks like a breakdown of the
evolution at the classical singularity. However, it turns out, in
exactly the same way as in the isotropic case, that we can evolve
through the singularity: the values of the wave function at the
classical singularity remain undetermined, but since they completely
drop out of the evolution equation they are not needed to find the
wave function at negative times $n_3$. The initial data chosen at
positive $n_3$, together with boundary data, uniquely determine the
wave function at positive and negative $n_3$ and the evolution through
the classical singularity is perfectly well-defined. Here we also used
the fact that matter Hamiltonians quantized with quantum geometrical
methods \cite{QSDV} are zero at the classical singularity such that
also the right hand side of (\ref{Evolve}) does not contribute values
of the wave function at the classical singularity. Again, it is
essential that the orientation of the triad provides us with twice as
many states as expected in a metric formulation. Thereby we naturally
obtain negative values of our internal time. Still, just using
variables which can take negative values is not enough since, most
importantly, the evolution must remain well-defined. This is a
non-trivial fact which in this case relies in part on the discreteness
of the triad minisuperspace. As already alluded to earlier, another
non-trivial fact (which is not present in the isotropic case) is that
the classical singularity lies in the interior of minisuperspace in
the densitized triad variables which we have to use in quantum
geometry, while it would be at the infinite boundary for, e.g.,
co-triad variables. It would be harder to see in general if a
singularity at the boundary of minisuperspace can be eliminated with
the present mechanism.

We also note that choosing a different factor ordering of the
constraint would lead to a singularity since some values of the wave
function could not be determined from the initial data but now would
not drop out completely and thus be necessary to determine other
values. The ordering we have to use here for a non-singular evolution,
corresponding to ordering the triads to the right, is the same as in
the isotropic case.

In order to discuss the boundary conditions we first transform the
evolution equation by introducing
\[
 t_{m_1,m_2,m_3}:=V_{2m_1,2m_2,2m_3}\; s_{2m_1,2m_2,2m_3}
\]
and defining
\begin{equation} \label{dm}
 d(m):=\left\{ \begin{array}{cl}\sqrt{|1+\case{1}{2}m^{-1}|}-
       \sqrt{|1-\case{1}{2}m^{-1}|} & \mbox{for }m\not=0 \\ 0 &
       \mbox{for }m=0 \end{array}\right.\,.
\end{equation}
Since the volume eigenvalues $V_{n_1,n_2,n_3}$ vanish for
$n_1n_2n_3=0$, the function $t$ has to fulfill
\begin{equation}\label{tCond}
 t_{0,m_2,m_3}=t_{m_1,0,m_3}=t_{m_1,m_2,0}=0
\end{equation}
for all $m_1,m_2,m_3$. In addition, $t$ also has to fulfill the gauge
invariance conditions (\ref{Ident}) such that we can restrict again to
non-negative $m_1,m_2$. Equation (\ref{Evolve}) then turns into
\begin{eqnarray}
 && d(m_2)(t_{m_1+1,m_2,m_3+1}-t_{m_1-1,m_2,m_3+1})+
 d(m_1)(t_{m_1,m_2+1,m_3+1}- t_{m_1,m_2-1,m_3+1})\nonumber\\
 &&+d(m_3)(t_{m_1-1,m_2-1,m_3}-t_{m_1+1,m_2-1,m_3}+
 t_{m_1+1,m_2+1,m_3}-t_{m_1-1,m_2+1,m_3}) \nonumber\\
 &&+d(m_2)(t_{m_1-1,m_2,m_3-1}-t_{m_1+1,m_2,m_3-1})+
 d(m_1)(t_{m_1,m_2-1,m_3-1}- t_{m_1,m_2+1,m_3-1})\nonumber\\
 &=& -\gamma^3\kappa\lP^2 \;\hat{\rho}_{\rm matter}(m_1,m_2,m_3)\;
 t_{m_1,m_2,m_3} \label{Evolvet}
\end{eqnarray}
where
\begin{equation}
 \hat{\rho}_{\rm matter}(m_1,m_2,m_3)=\hat{H}_{\rm
   matter}(m_1,m_2,m_3)/V_{m_1,m_2,m_3}
\end{equation}
for non-zero $m_1$, $m_2$, $m_3$ and zero otherwise, is the matter
energy density operator.

Starting with initial data at two slices of constant $m_3$ yields via
recurrence the combination
\[
 d(m_2)(t_{m_1+1,m_2,m_3+1}-t_{m_1-1,m_2,m_3+1})+
 d(m_1)(t_{m_1,m_2+1,m_3+1}- t_{m_1,m_2-1,m_3+1})
\]
of values of the wave function at the next step $m_3+1$. For this to
yield unique values $t_{m_1,m_2,m_3+1}$ we need to specify boundary
values at low values of $m_1$ and $m_2$. For $m_1=0$ we obtain the
values of
\[
 d(m_2)(t_{1,m_2,m_3+1}-t_{-1,m_2,m_3+1})=
 d(m_2)(t_{1,m_2,m_3+1}-t_{1,m_2,-m_3-1})
\]
from which we can find $t_{1,m_2,m_3}$ at negative $m_3$ provided that
the values $t_{1,m_2,m_3}$ for positive $m_3$ are given by boundary
conditions. The value $m_1=1$ in the recurrence relation yields
\[
 d(m_2)t_{2,m_2,m_3+1}+ d(1)(t_{1,m_2+1,m_3+1}- t_{1,m_2-1,m_3+1})
\]
(using $t_{0,m_2,m_3+1}=0$) which allows us to determine
$t_{2,m_2,m_3+1}$ uniquely. All other values of $t$ for $m_1>2$ are
fixed in the same way, and a similar analysis holds for varying
$m_2$. This shows that we need to specify the values
\begin{equation}
 t_{1,m_2,m_3} \mbox{ and }t_{m_1,1,m_3}\mbox{ for all }m_1,m_2,m_3>0
\end{equation}
as boundary data together with initial values at fixed $m_3$.

\begin{figure}
 \begin{center}
 \includegraphics[width=12cm,height=8cm]{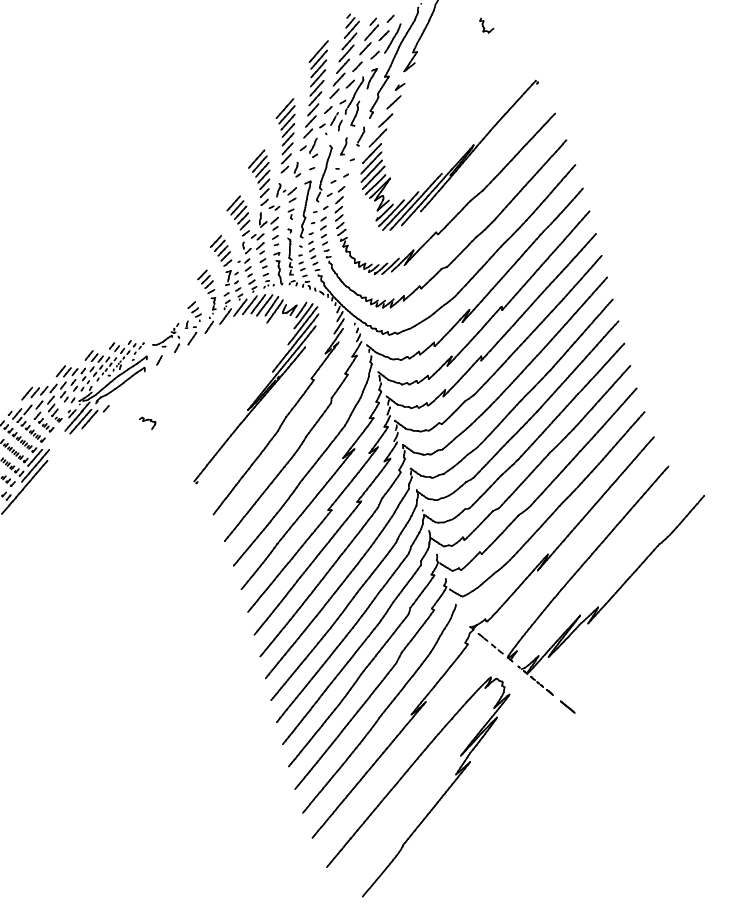}
 \end{center}
 \caption{Contour lines for a wave packet of the Bianchi I LRS model
  moving through the classical singularity (the diagonal line where
  the wave function drops to zero). The lower right boundary
  corresponds to the boundary $m=0$ of the minisuperspace, while the
  upper left boundary has been introduced for numerical purposes. (Not
  all values of the wave function are plotted; see the Appendix for
  further explanation.)}
 \label{Pack}
\end{figure}

\begin{figure}
 \begin{center}
 \includegraphics[width=12cm,height=8cm]{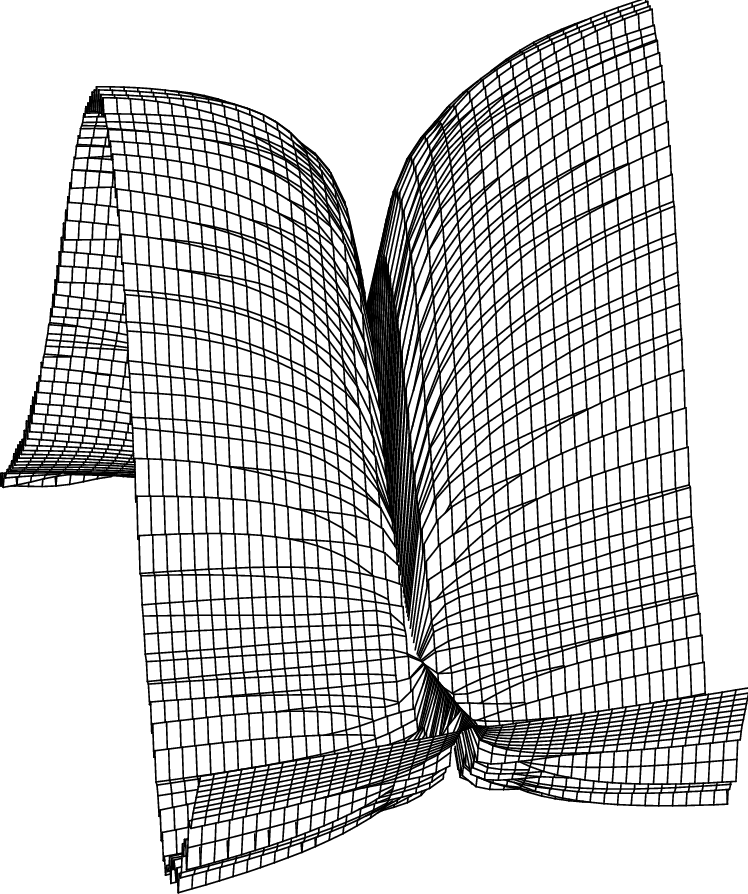}
\end{center}
 \caption{Closer look at the classical singularity. Lattice points are
 values of the discrete wave function $t_{mn}$.}
 \label{PackClose}
\end{figure}

As for the initial values, one would expect to need the wave function
at two slices of constant $m_3$ because the evolution equation is of
second order. However, by definition $t_{m_1,m_2,m_3}$ has to vanish
for $m_3=0$ which is equivalent to the fact that $s_{n_1,n_2,0}$ drops
out of the evolution equation. Thus, arbitrary initial values at fixed
$m^{(0)}_3$ and $m^{(0)}_3+1$ will not be consistent with this
condition and the evolution to $m_3=0$ will yield linear relations for
the initial data. Consequently, only the values at one slice of fixed
$m^{(0)}_3$ are free and the number of initial conditions is only half
the expected amount, as in the isotropic case. Unlike the isotropic
situation, however, this does not suffice to specify the gravitational
part of the wave function uniquely up to norm. Further conditions can
be expected to arise from the requirement of a pre-classical wave
function at large volume \cite{DynIn} where the wave function should
not oscillate on Planck scales. This also restricts the original wave
function $s_{n_1,n_2,n_3}$ at odd values of $n_I$, which are not
determined by $t_{m_1,m_2,m_3}$. Such an analysis, however, is more
complicated for a partial difference equation and will be pursued
elsewhere. Figs.\ \ref{Pack} and \ref{PackClose} show two different
impressions of a solution to the Hamiltonian constraint of the Bianchi
I LRS model which represents a wave packet moving through the
classical singularity. Details of this model and its classical and
Wheeler--DeWitt behavior, as well as more informations about the
figures, can be found in the Appendix.

\subsection{Flat Space-Time}

It is usually hard to find exact solutions of a partial difference
equation as (\ref{Evolvet}) with non-constant coefficients, even if it
is linear. In our case it is easy to see that $t_{m_1,m_2,m_3}=m_1$,
$t_{m_1,m_2,m_3}=m_2$, and $t_{m_1,m_2,m_3}=m_3$ are exact solutions
in the vacuum case. However, they do not satisfy the conditions
(\ref{tCond}) and thus are not allowed.

An approximate solution at large volume can be seen to be of the form
$t_{m_1,m_2,m_3}=m_1m_2m_3$. It does fulfill the conditions
(\ref{tCond}), but this is not relevant since the solution is valid
only approximately at large $m_1$, $m_2$, $m_3$. To see that this is
an approximate solution we expand $d(m)=\frac{1}{2}m^{-1}+O(m^{-2})$
such that the left hand side of (\ref{Evolvet}) becomes the constant
$6$ up to terms of the order $m_I^{-1}$. This corresponds to a
negative matter energy density $\rho_{\rm matter}\sim -6
\gamma^{-3}\kappa^{-1}\lP^{-2} V^{-2}$ which decreases like the
inverse second power of the volume and thus would be negligible
compared to any other standard matter component at large volume. In
fact, the solution $t_{m_1,m_2,m_3}=m_1m_2m_3$, which corresponds to
$s_{n_1,n_2,n_3}\propto \sgn(n_1n_2n_3)\sqrt{|n_1n_2n_3|}\propto
\sgn(n_1n_2n_3)V_{n_1,n_2,n_3}$, represents flat space in the sense
that all terms in the constraint $c_1c_2a_3+c_2c_3a_1+c_3c_1a_3=0$
which follows from (\ref{H}) vanish separately. At large volume a
quantization of $a_3=\sqrt{|p^1p^2/p^3|}\sgn(p^3)$ must have
approximate eigenvalues of the form $\sqrt{|n_1n_2/n_3|}\sgn(n_3)$
such that in the triad representation we have for our approximate
solution $(\hat{a}_3s)_{n_1,n_2,n_3}\propto n_1n_2$. In the connection
representation, this corresponds to the state
\[
 \hat{a}_3|s\rangle= \sum_{n_1,n_2,n_3}n_1n_2\;
 |n_1,n_2,n_3\rangle\propto\tilde\delta(c_1)\tilde\delta(c_2) \tilde0(c_3)
\]
where $\tilde\delta(c):=\sum_j (2j+1)\chi_j(c)$ denotes the
$\delta$-distribution on $SU(2)$ and also
$\tilde0(c):=\sum_j\zeta_j(c)$ is to be interpreted in a distributional
sense.  Therefore, one would expect that $\hat{a}_3|s\rangle$ is
annihilated by a quantization of $c_1c_2$ as a multiplication
operator, which appears in the constraint. However, as already seen,
the annihilation is not complete which leads to the matter
contribution decreasing as the inverse second power of the volume. To
understand this fact we have to take into account that the
configuration space (for a single diagonal component) differs from
$SU(2)$ precisely in the singularity structure of the point $c=0$
which is a fixed point of the gauge transformations. Correspondingly,
the state space for a single diagonal component is not given simply by
gauge invariant functions on $SU(2)$, i.e., only the characters
$\chi_j$, but by all possible functions generated by $\chi_j$ and
$\zeta_j$. Since $\zeta_j$ is not regular in $c=0$ for any $j$, an
ordinary $\delta$-distribution on this space of states cannot be
well-defined. Instead, the distribution $\tilde\delta$ introduced
above appears which can be regarded as a regularization of the
ill-defined naive $\delta$-distribution.

For regular test functions $f(c)$, $\tilde\delta$ in fact has the
expected action, as can be checked in the usual way:
\begin{eqnarray*}
 (2\pi)^{-1}\int\tilde\delta(c)f(c)\sin^2(\case{1}{2}c)\md c &=&
 (2\pi)^{-1}\sum_j\int(2j+1)
 \sin\left((j+\case{1}{2})c\right)\sin(\case{1}{2}c)f(c)\md c\\ &=&
 \pi^{-1}\sum_j\int_0^{4\pi}\cos\left((j+\case{1}{2})c\right)\frac{\md}{\md
 c} \left(f(c)\sin(\case{1}{2}c)\right) \md c\\ &=&
 2\pi^{-1}\sum_{n\in\N}\int_0^{2\pi}\cos (nc)(f'(2c)\sin
 c+\case{1}{2}f(2c)\cos c)\md c\\ 
 &=& 2(f'(2c)\sin c+\case{1}{2}f(2c)\cos c)|_{c=0}\\
 && - \pi^{-1}\int_0^{2\pi}(f'(2c)\sin
 c+\case{1}{2}f(2c)\cos c)\md c\\ &=& f(0)
\end{eqnarray*}
where we used the $U(1)$ $\delta$-distribution
$\delta(t)=(2\pi)^{-1}\sum_{n\in\Z}
e^{int}=(2\pi)^{-1}(2\sum_{n\in\N}\cos (nt)+1)$ and periodicity of $f$
and $f'$. If $f$ is not regular in $c=0$, however, i.e., of the form
$f(c)=g(c)\sin(\case{1}{2}c)^{-1}$ with a periodic function $g$ which
is regular and non-zero in $c=0$, we have
\begin{eqnarray*}
 (2\pi)^{-1}\int\tilde\delta(c)f(c)\sin^2(\case{1}{2}c)\md c &=&
 (2\pi)^{-1}\sum_j\int (2j+1)\sin\left((j+\case{1}{2})c\right) g(c)\md
 c\\ &=& 2\pi^{-1}\sum_{n\in\N}\int_0^{2\pi} \cos (nc)g'(c)\md c= 2g'(0)\,.
\end{eqnarray*}
Thus, the regularized $\delta$-distribution is
\begin{equation}
 \tilde\delta[f]=2\frac{\md}{\md
   c}\left(\sin(\case{1}{2}c)f(c)\right)|_{c=0}=
 \left\{\begin{array}{cl} f(0) & \mbox{if $f$ regular in $c=0$}\\
     2g'(0) & \mbox{if }
     f(c)=g(c)\sin(\case{1}{2}c)^{-1}\end{array}\right.\,.
\end{equation}

Since the $\delta$-distribution is regularized in order to take into
account non-regular test functions, it is not annihilated when
multiplying with $\sin(\case{1}{2}c)$ or $c$. Instead, we have
\begin{eqnarray*}
 (2\pi)^{-1}\int
 \sin(\case{1}{2}c)\tilde\delta(c)f(c)\sin^2(\case{1}{2}c)\md c &=&
 2\pi^{-1}\int_0^{2\pi}\sum_{n\in\N} \cos (nc)(f'(2c)\sin^2c+f(2c)\sin
 c\cos c)\md c\\ &=& 0 \quad\mbox{if $f$ regular}
\end{eqnarray*}
but
\begin{eqnarray*}
 (2\pi)^{-1}\int
 \sin(\case{1}{2}c)\tilde\delta(c)f(c)\sin^2(\case{1}{2}c)\md c &=&
 (2\pi)^{-1}\int\tilde\delta(c)g(c)\sin^2(\case{1}{2}c)\md c\\ &=&
 g(0) \quad\mbox{otherwise.}
\end{eqnarray*}
In fact, it is easy to check that we obtain the distribution $\tilde0$
defined above when we multiply with $\sin(\case{1}{2}c)$:
\begin{eqnarray*}
 \tilde0[f] &=& (2\pi)^{-1}\int\tilde0(c)f(c)\sin^2(\case{1}{2}c)\md c
 = (2\pi)^{-1}\sum_{n\in\N}\int \cos(\case{1}{2}nc)
 f(c)\sin(\case{1}{2}c)\md c\\ &=& (f(2c)\sin
 c)|_{c=0}=\left\{\begin{array}{cl} 0 & \mbox{if $f$ regular}\\ g(0) &
 \mbox{if }f(c)=g(c)\sin(\case{1}{2}c)^{-1}
   \end{array}\right.\,.
\end{eqnarray*}
Thus, $\sin(\case{1}{2}c)\tilde\delta=\tilde0$ and we have to multiply
a second time to get zero, $\sin(\case{1}{2}c)\tilde0=0$. 

In the isotropic model we have the same structure of the singularity
$c=0$ in the configuration space, but there the constraint is
quadratic in the single gauge invariant connection component $c$ which
annihilates the regularized $\delta$-distribution. Therefore, there is
no remaining matter contribution in the isotropic case. This
demonstrates how certain features of a theory can be lost when
additional symmetries are introduced, and also that working with
gauge-invariant $SU(2)$-states (\ref{chi}) and (\ref{zeta}) can lead
to results different from those obtained with $U(1)$-states $e^{inc}$
which would appear in a gauge-fixed theory. In this example, however,
the physical relevance is not clear. Furthermore, it is an effect of
the singularity structure in $c_I=0$ which we can probe here since we
are looking for solutions of flat space-time which are $c_I=0$
classically. For other solutions the set $c_I=0$, which is of measure
zero in the minisuperspace, is not expected to contribute
significantly. However, such solutions corresponding to the classical
Kasner behavior are harder to find even approximately, and one has to
resort to a numerical analysis. Some difficulties to be faced are
discussed in the Appendix.

\section{Discussion}

In this paper we developed loop quantum cosmological techniques for
homogeneous cosmological models. The calculations have been simplified
considerably by diagonalizing the canonical degrees of freedom, a
process which retains all the relevant physical information. Most
importantly, the volume operator is simpler and it is possible to find
its spectrum explicitly. Having a simple volume operator then also
leads to explicit expressions such as inverse metric operators and the
Hamiltonian constraint operator which are necessary to study the
classical singularity. Extending the results for isotropic models, we
now have practical techniques for a larger class of systems given by
all Bianchi class A models and their locally rotationally symmetric
submodels (which have a single rotational axis). Those models are
dynamically non-trivial even in the vacuum case such that they allow
us to study quantum geometry effects independently of the matter
coupling.

Here we mainly applied the techniques to the Hamiltonian constraint of
the Bianchi I model. It turned out that the classical singularity is
absent in loop quantum cosmology in very much the same way as in the
isotropic case. In particular, it is the same factor ordering of the
constraint operator which is selected by the requirement of a
non-singular evolution. New features compared to the isotropic case
also emerged: it has been seen to be essential that quantum geometry
is based on densitized triad variables, and the issue of initial
conditions is more complex. One has to choose not just initial
conditions but also boundary conditions on minisuperspace, and the
issue of pre-classicality is more complicated to analyze. It may turn
out that pre-classical wave functions (up to norm) fulfilling the
consistency conditions are not unique, but still the freedom is
reduced compared to the Wheeler--DeWitt approach.

The analysis has also been simplified by choosing a diagonal component
of the triad as internal time, rather than the volume. Even though the
volume spectrum is known explicitly, its use as internal time would
lead to a more complicated evolution equation since it is not equally
spaced. For the Bianchi I model with its classical Kasner behavior, a
diagonal component as internal time is justified, and it is well
suited to study the approach to the classical singularity. In other
Bianchi models the behavior is certainly more complicated: one expects
a succession of different Kasner epochs. While in all those epochs the
triad components $p^I$ are all decreasing toward the singularity, it
is not always clear what happens during transitions between different
epochs. If the behavior of the triad components remains monotonic
during the transitions, one can use one of them as internal time and
follow the evolution. Note that the result of an absence of a
singularity will hold true in any case since it is a statement about
the possibility of extending the wave function throughout all of
minisuperspace uniquely; choosing an internal time is then only
necessary to develop an evolution picture.

Particularly interesting is the Bianchi IX model which has been
suspected to behave chaotically. Here one can check if the modified
approach to classical singularities in quantum geometry, which has
been observed in the isotropic case in \cite{Inflation} and been seen
to lead to inflation, implies a different behavior. In isotropic
models, the origin of a modified approach comes from quantum geometry
effects in the matter Hamiltonian while the gravitational part of the
effective classical constraint equation has been left unmodified in a
first approximation.  In Bianchi models other than type I one expects
a similar modification even in the vacuum case because the spin
connection (\ref{SpinConn}) is non-zero and depends on triad
components. This has to be dealt with properly when quantizing the
Hamiltonian constraint and opens the door for modifications in
effective classical equations of motion. While this does not modify
the kinetic part (i.e., the terms containing time derivatives in
(\ref{Ha})) of the Hamiltonian constraint in a first approximation, it
leads to a modified potential containing the spin connection at small
volume. One can then expect that the series of transitions between
Kasner epochs, which can be seen as reflections of the scale factors
on the potential well, will change since the potential well breaks
down if the volume is small enough. This may stop the reflections
altogether and lead to a non-chaotic behavior. Then, a single Kasner
epoch would be responsible for the transition through the classical
singularity, which would not look much different from the Bianchi I
behavior close to the singularity. These issues will be discussed in
more detail elsewhere.

The Bianchi IX model also provides further clues for the full
theory. It has been speculated that the approach to inhomogeneous
singularities is described by the BKL scenario \cite{BKL} according to
which different points on a space-like slice decouple from each other
close to a singularity (technically, only time derivatives remain in
an asymptotic expansion of Einstein's equations close to a
singularity). If this scenario is true and can be used in the quantum
theory, it suggests that a non-singular Bianchi IX behavior implies
non-singular behavior in the full theory. However, since even the
classical validity of the BKL scenario is far from being established,
such a conclusion would be premature at present.

An alternative route is to study less symmetric models and see if what
we have learned from more symmetric ones remains true. In this paper
we presented the first step, going from isotropic models to
anisotropic but still homogeneous ones. It turned out that essential
results did not change, in particular the evolution remains
non-singular. Even though the mechanism is similar to that in
isotropic models, there are non-trivial aspects which cannot be seen
in an isotropic model. It could well have happened that there is a
singularity in homogeneous models but not in isotropic ones, e.g., if
one used a theory based on the co-triad rather than the densitized
triad as fundamental variable. That quantum geometry so far has
provided precisely those properties one would need to have
non-singular behavior is encouraging, but it also indicates that there
may still be further surprises when we go one step further by studying
inhomogeneous midisuperspace models.

\section*{Acknowledgements}

The author is grateful to A.~Ashtekar, G.~Date, C.~Fleischhack,
H.~Morales-T\'ecotl, and K.~Vandersloot for discussions and
comments. This work was supported in part by NSF grant PHY00-90091 and
the Eberly research funds of Penn State.

\section*{Appendix}

\begin{appendix}
\renewcommand{\theequation}{\thesection.\arabic{equation}}
\setcounter{equation}{0}

\section{The Bianchi I LRS Model}

Probably the most simple anisotropic model is the Bianchi I LRS model
without matter, which is defined as the subclass of Bianchi I
geometries admitting one spatial rotational symmetry axis. Therefore,
two of the diagonal components of the connection as well as of the
triad, e.g.\ the first two for definitiveness, have to equal each
other and only two degrees of freedom are left which we choose to be
$(c,p_c)$ and $(A,p_A)$ embedded into the Bianchi I model by
\[
 (A,c)\mapsto (c_1,c_2,c_3)=(A,A,c)\qquad,\qquad (p_A,p_c)\mapsto
 (p^1,p^2,p^3)=(p_A,p_A,p_c)\,.
\]
The symplectic structure derived from the embedding follows from
(\ref{symp}):
\[
 \{A,p_A\}=\case{1}{2}\gamma\kappa\qquad,\qquad\{c,p_c\}=\gamma\kappa\,.
\]

For the LRS model the constraint (\ref{H}) reduces to
\[
 H=-2\gamma^{-2}\kappa^{-1} |p_c|^{-\frac{1}{2}} A(2cp_c+Ap_A)\propto
 Ap_A+2cp_c
\]
which implies the equations of motion
\[
 \dot{p}_A=\case{1}{2}p_A\quad,\quad \dot{A}=-\case{1}{2}A\quad,\quad
 \dot{p}_c=2p_c\quad,\quad \dot{c}=-2c
\]
solved by
\[
 p_A(\tau)\propto\sqrt{\tau}\quad,\quad A(\tau)\propto
 \tau^{-\frac{1}{2}}\quad,\quad p_c(\tau)\propto \tau^2\quad,\quad
 c(\tau)\propto \tau^{-2}\,.
\]
Using $p_c$ as internal time we can write the evolution of $p_A$ as
$p_A(p_c)\propto p_c^{\frac{1}{4}}$, which coincides with the Kasner
behavior restricted to the LRS case: If two of the Kasner coefficients
are equal, we have $2\alpha_1+\alpha_3=1=2\alpha_1^2+\alpha_3^2$
implying $\alpha_1=\frac{2}{3}$ and $\alpha_3=-\frac{1}{3}$. In the
Kasner time coordinate $t$ we then have
$p_A\propto t^{\frac{1}{3}}$, $p_c\propto t^{\frac{4}{3}}$
according to (\ref{pa}), which also leads to $p_A(p_c)\propto
p_c^{\frac{1}{4}}$.

\begin{figure}
 \begin{center}
 \includegraphics[width=12cm,height=8cm]{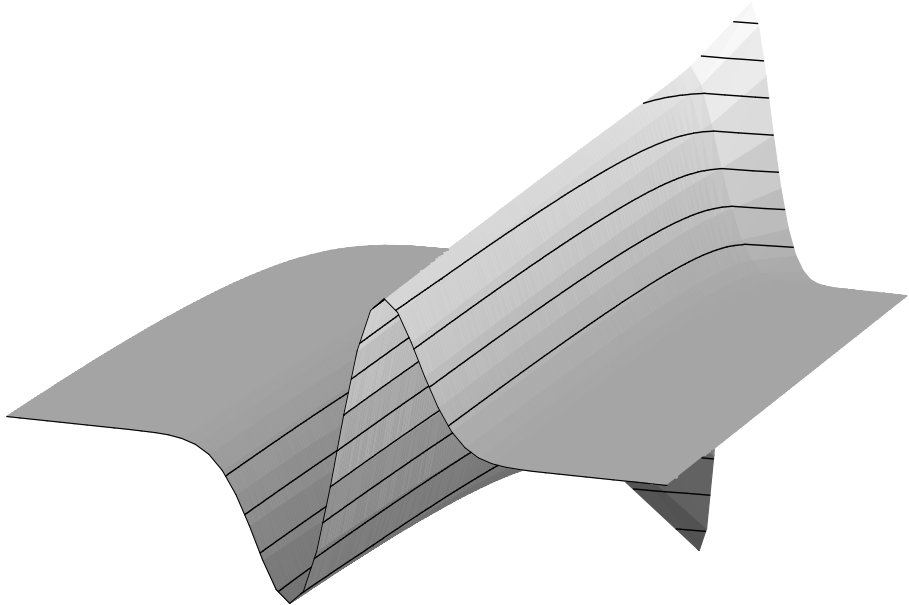}
\end{center}
 \caption{A solution to the Wheeler--DeWitt equation of the Bianchi I
 LRS model. When the wave packet approaches the classical singularity
 (upper right corner), its amplitude diverges. The minisuperspace
 boundary $p_A=0$ is at the right.}  \label{LRScont}
\end{figure}

The Wheeler--DeWitt equation in the triad representation for the
Hamiltonian $Ap_A+2cp_c$ can be written in the form
\[
 \left(\case{1}{2}p_c^{-1}\frac{\partial}{\partial
 p_A}p_A+2\frac{\partial}{\partial p_c}\right)\tilde{\psi}(p_A,p_c)=0
\]
where a factor of $p_c$ has been absorbed in the wave function
$\tilde{\psi}(p_A,p_c):=p_c\psi(p_A,p_c)$. Fig.~\ref{LRScont} shows a
solution which represents a wave packet following the classical
trajectory $p_A(p_c)\propto p_c^{\frac{1}{4}}$ in minisuperspace,
hitting the classical singularity. 

In the loop quantization we have only two integer labels $m$ and $n$,
where $m$ represents the value of the two equal triad components $p_A$
and $n$ the third one, $p_c$. Residual gauge transformations can now be
seen to change the sign of $m$ such that we can get unique
representatives of gauge equivalence classes by considering only
non-negative values of $m$. The label $n$, on the other hand, is
allowed to take any integer value and thus is best suited to describe
the transition through the classical singularity. We will use it as
internal time, the sign of which, $\sgn(n)$, is again the orientation.

For the sake of simplicity we write down directly the evolution
equation for a wave function $t_{m,n}$ in the form (\ref{Evolvet}) which
now becomes
\begin{eqnarray*}
 && 2d_2(m) (t_{n+1,m+1}-t_{n+1,m-1})+ d(n)
 (t_{n,m-2}-2t_{n,m}+t_{n,m+2})\\
 &&+ 2d_2(m) (t_{n-1,m-1}-t_{n-1,m+1})\\
 &=&
 2d_2(m)\tilde{t}_{n+1,m}+
 d(n)(\tilde{t}_{n,m+1}-\tilde{t}_{n,m-1})-2d_2(m)\tilde{t}_{n-1,m}=0\,.
\end{eqnarray*}
Here, we defined $\tilde{t}_{n,m}:=t_{n,m+1}-t_{n,m-1}$ and
$d_2(m):=m^{-1}$ for $m\not=0$ and $d_2(0):=0$ (it replaces $d(m)$ of
(\ref{dm}) since now two labels in the volume eigenvalues are equal to
each other).

We now have the evolution equation in the form of a partial difference
equation in two independent variables, which is straightforward to
implement numerically. It is, however, a non-trivial problem to find
numerical solutions which are close to pre-classical ones, i.e., which
vary significantly only on large scales. We have many parameters for a
solution, initial values at one $n$-slice and boundary values at small
$m$. To find a solution which satisfies the consistency condition
($\tilde{t}_{0,m}=0$), we can choose the initial values at the
$n$-slice and the boundary values freely. A complete initial value
problem would be specified by also choosing values at the next
$n$-slice since the evolution equation is of second order in $n$. We
can find the values of the second slice by first leaving them as free
parameters, solving the evolution equation up to $n=0$ where we obtain
$\tilde{t}_{0,m}$ as linear functions of the free initial values, and
solving the set of linear equations $\tilde{t}_{0,m}=0$. The result is
a solution which by construction satisfies the consistency
condition. In practice, we have to face the following problems when we
want to find a numerical solution which is also pre-classical:
\begin{itemize}
\item There is no guarantee that the computed consistent initial data
  at the second $n$-slice are close to the ones chosen at the first
  slice (which would be necessary for pre-classicality). One can try
  to find such a solution by picking appropriate initial and boundary
  values, but it is complicated by the large amount of initial and
  boundary parameters we have to choose.
\item For numerical purposes we have to introduce a second boundary at
  large $m$, which can introduce artificial boundary effects.
\item The initial values should represent a pre-classical function
  such that differences appearing in the evolution equation are
  small. This requires a large lattice in the $m$-direction which
  increases the numerical effort to evolve and to solve the system of
  linear equations in order to find the initial values at the second
  $n$-slice. Furthermore, more $m$-values increase the likelihood of
  rounding errors adding up.
\item Close to classical singularities the wave packet can move
  rapidly on the lattice (corresponding to large extrinsic
  curvature). If this happens even between two consecutive $n$-slices,
  differences in the $n$-direction will become large which can lead to
  contributions increasing exponentially toward the classical
  singularity. While imposing the consistency conditions exactly would
  eliminate those contributions, they will always be present in a
  numerical analysis and can even lead to wrong initial values due to
  rounding errors.
\end{itemize}

In order to show an illustration of a possible pre-classical solution,
Figs.~\ref{Pack} and \ref{PackClose} have been obtained as follows:
First, the evolution equation shows that for large $n$ we have
$\tilde{t}_{n+1,m}\simeq\tilde{t}_{n-1,m}$, provided that the
$m$-differences of $\tilde{t}$ are small. Therefore, we can suppress
possible oscillations in the $n$-direction, coming from consistent
initial values at the second $n$-slice not close to the one on the
first slice, by plotting only every second $n$-value. Secondly, in
order to avoid a rapid movement of the wave packet in the
$m$-direction, the lattice has been stretched in the $n$-direction by
using the Hamiltonian $H_k=Ap_A+2kcp_c$. While the original value
$k=1$ leads to a rapid movement at small $n$ which implies rounding
errors due to exponentially increasing contributions and an
unacceptable numerical solution, this turns out not to pose a problem
for $k\geq4$ ($k=4$ in the figures) since the movement would happen
between $n=0$ and $n=1$ which is not visible.

One could think that it is even impossible to find a pre-classical
wave function reflecting the fact that DeWitt's initial condition,
which is close to the result of the dynamical initial condition, is
ill-posed in this model (i.e., its only solution would be zero). This,
however, is not the case for two reasons: first, the notion of
pre-classicality allows some tolerance since a discrete wave function
can only be required to be close to a smooth one, but not exactly
smooth; and secondly, the discreteness leads to a different behavior
of solutions at small $n$ such that a DeWitt-like initial condition is
in fact well-posed (see \cite{Scalar}).

\end{appendix}


\end{document}